\documentclass[12pt]{article}
\usepackage{
fullpage}
\usepackage{wrapfig}
\usepackage{amsmath}
\usepackage{graphicx,psfrag,epsf}
\usepackage{enumerate}
\usepackage{natbib}
\usepackage{amsmath,amsthm,amsfonts,epsfig,amssymb,natbib,eucal,eufrak}
\usepackage{booktabs}
\usepackage{multirow}
\usepackage{siunitx}
\usepackage{qtree}
\usepackage{color}

\usepackage{float}
\floatstyle{plaintop}
\restylefloat{table}

\newcommand{\blind}{0}

\begin{document}

\def\spacingset#1{\renewcommand{\baselinestretch}%
{#1}\small\normalsize} \spacingset{1}


\if0\blind
{
  \title{\bf Nested Group Testing Procedures for Screening}
  \author{Yaakov Malinovsky\thanks{ Department of Mathematics and Statistics, University of Maryland, Baltimore County, Baltimore, MD 21250, USA}\,\,\,
    and\,
    Paul S. Albert\thanks{
    Biostatistics Branch, Division of Cancer Epidemiology and Genetics, National Cancer Institute, Rockville, MD 20850, USA.
    The work was supported by the National Cancer Institute Intramural Program.}
}
  \maketitle
} \fi


\bigskip
\begin{abstract}
This article reviews a class of adaptive group testing procedures that operate under a probabilistic model assumption as follows.
Consider a set of $N$ items, where item $i$ has the probability $p$ ($p_i$ in the generalized group testing) to be defective, and the probability $1-p$ to be non-defective independent from the other items. A group test applied to any subset of size $n$ is a binary test with two possible outcomes, positive or negative.
The outcome is negative if all $n$ items are non-defective, whereas the outcome is positive if at least one item among the $n$ items is defective.  The goal is complete identification of all $N$ items with the minimum expected number of tests.

\end{abstract}

\noindent%


\section{Introduction}
In the last few months of the COVID-19 pandemic, the mostly-forgotten practice of group testing has been raised again in many countries as an efficient
method for addressing an epidemic while facing restrictions of time and resources.
During this very short period, numerous publications and reports have appeared in both scientific and non-scientific journals.
The reader can easily see them, for example, in the {\it Washington Post, NY Times, Scientific American, Science Advances, medRxiv, bioRxiv, ArXiv}, and elsewhere.

The story goes back to 1943, when Robert Dorfman published a manuscript where he introduced the concept of group testing
in response to the need to administer syphilis tests to millions of individuals drafted into the U.S. Army during World War II \citep{Dorfman1943}.
A nice description of the \cite{Dorfman1943} procedure is given by \cite{F1950}:
 {\it "A large number, $N$, of people are subject to a blood test. This can be administered in two ways. (i) Each person is tested separately. In this case $N$ tests are required. (ii) The blood samples of $k$ people can be pooled and analyzed together. If the test is negative, this one test suffices for the $k$ people. If the test is positive, each of the $k$ persons must be tested separately, and all $k+1$ tests are required for the $k$ people. Assume the probability $p$ that the test is positive is the same for all  and that people are stochastically independent.}"\\
Procedure $(ii)$ is commonly referred to as the Dorfman (D) two-stage group testing (GT) procedure.

In general, the above setting assumes a probabilistic model where there are $N$ individuals to be tested, test outcomes are independent, and each individual has the same probability $p$ to be infected (Binomial model). In this article, we will discuss different
GT procedures under this model, where comparisons are done based on the expected number of tests.
The fundamental result by Peter Ungar \citep{U1960} shows that if $p>p_{u}=(3-\sqrt{5})/2\approx 0.38$, then individual testing is the optimal GT procedure, and if $p<p_u$, then it is not optimal.
However, it is important to note that despite 80 years' worth of research effort, the optimal procedure is yet unknown for $p<p_{u}$ and a general $N$.

This article deals with nested GT procedures, of which the \cite{Dorfman1943} procedure is a member.
A nested algorithm has the property that if a positive subset $I$ is identified, the next subset $I_1$ that we will test is a proper subset of $I$--that is, $I_1\subset I$. This natural class of GT procedures was defined by \cite{SG1959} and \cite{S1960}. Procedure $D$ belongs to this class with the restriction that up to two stages are required.

\section{Nested GT Procedures: common $p$ case}
Recall that we prefer procedure/design $A$ over procedure/design $B$ if $E_{A}\leq E_{B}$, where $E_{}$ is the corresponding expected number of tests.

\subsection{Dorfman and modified Dorfman procedures}
In a group of size $k\geq 2,$ the total number of tests is $1$ with probability $q^k$ ($q=1-p$) and $k+1$ with probability $1-q^k$. Therefore, the expected number of tests per person in Procedure D is
${\displaystyle E_{D}\left(k, p\right)=1-q^{k}+\frac{1}{k},\,\text{for}\,\,k\geq 2}$; and equals 1 otherwise.

Dorfman numerically found the optimum group size $k$ when assuming that a population is large (infinite) and $p$ is fixed.
For example, if $p=0.01$, then the optimal group size is $11$.
Specifically, this means that for testing $N=999,999$ individuals, we need only $195,571$ tests in expectation.
\cite{S1978} showed that an optimal value of $k$, $k^{*}_{D}(p)$ is a non-increasing function of $p$, which is $1$ for $p>1-1/3^{1/3}\approx 0.31$ and otherwise is either $1+[p^{-1/2}]$ or $2+[p^{-1/2}]$, where $[x]$ denotes the integer part of $x$.

There is a logical inconsistency in Procedure {\it D}.
It is clear that any "reasonable" group testing plan should satisfy the following property:
"A test is not performed if its outcome can be inferred from previous test results" \citep{U1960}.
Procedure $D$ does not satisfy this property
since if the group is positive and all but the last person are negative, the last person is still tested.
The modified Dorfman procedure, which we define as $D^{\prime}$, would not test the last individual in that case \citep{SG1959}. Also, Procedure $D^{\prime}$  will be preferred over individual testing if $p$ is below Ungat's cut-off point \citep{MA2019}.
For $k\geq 2,$  the total number of tests is $1$ with probability $q^k$, $k$ with probability $q^{k-1}(1-q)$,
and $k+1$ with probability $1-q^{k-1}$. Therefore, the expected number of tests per person in a group of size $k$ under Procedure $D^{\prime}$: ${\displaystyle E_{{D^\prime}}\left(k, p\right)=1-q^k+1/k-(1/k)(1-q)q^{k-1}}.$
\cite{PE1978} showed that an optimal value $k^{*}_{D^{\prime}}$ is the smallest $k$ value that satisfies
$E_{D^{\prime}}\left(k, p\right)\leq E_{D^{\prime}}\left(k-1, p\right)\,\,\,\,\text{and}\,\,\,\,E_{D^{\prime}}\left(k, p\right)< E_{D^{\prime}}\left(k+1, p\right)
$. It was conjectured and empirically verified in \cite{MA2019} that
the optimal group size $k^{*}_{D^{\prime}}(p)$ is equal to either $\displaystyle \lfloor p^{-1/2}\rfloor$ or $\displaystyle \lceil p^{-1/2}\rceil$.

\subsection{Sterrett sequential procedure}
\cite{S1957} realized that one can improve the efficiency of the GT procedure by a sequential modification of Procedure $D^{\prime}$.
If in the first stage of Procedure $D^\prime$ a group is positive, then
individuals are tested one-by-one until the first positive individual is identified, or until all but the last person are negative; in the latter case, the positivity of the last individual follows from the positivity of the group.
Otherwise, if the first individual identified as positive is not the last in the group, then the first stage of Procedure $D^\prime$ is applied to the remaining (nonidentified) individuals.
This process is repeated until all individuals are identified.
A simple closed-form expression for the expected number of tests per person was provided by \cite{SG1959}:
${\displaystyle
E_{S}\left(k, p\right)=   \frac{1}{k}\left[2k-(k-2)q-\frac{1-q^{k+1}}{1-q}\right]}.$
It was conjectured and empirically verified in \cite{MA2019} that for $ 0<p<p_U$
the optimal group size $k^{*}_{D^{\prime}}(p)$ is
equal to $\displaystyle \lfloor \sqrt{2/p}\rfloor$ or $\displaystyle \lfloor \sqrt{2/p}\rfloor +1$ or $\displaystyle \lfloor \sqrt{2/p}\rfloor +2$.
Using above conjectures one can verify that ${\displaystyle \lim_{p\downarrow 0}\frac{E_{D}\left(k^*_{D}, p\right)}{E_{S}\left(k^*_{S}, p\right)}=\sqrt{2}.}$ Some extensions of the Sterrett (S) procedure were presented in \cite{J1991}.
\bigskip

\noindent
{\bf Finite Versus Infinite Population}
\smallskip

\noindent
A finite population of size $N$ is not necessarily divisible by $k$.
Therefore, for a finite population of size $N$ and a given Procedure $A\in \left\{D, D^{\prime}, S\right\}$, we have to solve the following optimization problem:
find the optimal {\it partition}
$\displaystyle \left\{n_1,\ldots,n_I\right\}$ with $n_1+\ldots+n_{I}=N$ for some $I\in\left\{1,\ldots,N\right\}$ such that $E_{A}\left(k,\, p\right)$  is minimal.
A common method to solve such an optimization problem is dynamic programming (DP) \citep{SG1959}.
It was conjectured by \cite{LS1972} for that Procedure $D$, the optimal partition subgroup sizes differ at most by one unit. \cite{G1985} proved a similar result for Procedure $D^{\prime}$, and \cite{MA2019} for Procedure $S$.

\subsection{Hierarchical and nested procedures}
Procedures $D, D^{\prime}$, and $S$ fit into a larger class of procedures discussed below.
\medskip

\noindent
{\bf An Optimal Hierarchical Procedure}
\medskip

The hierarchical class procedure was introduced by \cite{SG1959}
and defined as follows (see also \cite{HPE1981}):
A procedure is in the hierarchical class (HC) if two units are only tested together in a group if they have an identical test history, i.e. if each previous group test contains either both of them or none of them.

It follows from this definition that a procedure in the HC is similar to the multistage Dorfman procedure.
An optimal hierarchical procedure was obtained by \cite{SG1959} as a dynamic programming algorithm with computational cost $O(N^2)$, which they called Procedure $R_3$. This was recently computationally improved by \cite{Z2017} (see also \cite{M2019} for a discussion).
\bigskip

\noindent
{\bf An Optimal Nested Procedure}
\medskip

\noindent
This class of GT procedures was defined by \cite{SG1959} and \cite{S1960, S1967}
A nested procedure requires that between any two successive tests $n$ units not yet classified have to be separated into only (at most) two sets. One set of size $m\geq 0$, called the ``defective set," is known to contain at least one defective unit if $m\geq 1$ (it is not known which ones are defective or exactly how many there are). The other set of size $n-m\geq 0$ is called the ``binomial set" because we have no knowledge about it other than the original binomial assumption.
Either of these two sets can be empty in the course of experimentation; both are empty at termination.

The number of potential nested group testing algorithms is astronomical. For example, if $N=5$, then there are  $235,200$ possible algorithms \citep{MS1977}.
\cite{SG1959} overcame this problem by proposing a DP algorithm that
finds the optimal nested algorithm, which Sobel and Groll termed ``Procedure $R_1$."
There was a large research effort to reduce the $O(N^3)$ computational complexity of the original proposed algorithm \citep{S1960,KS1971,H1976a, YH1990}.
\cite{ZP2016} provided
an asymptotic analysis of the optimal nested procedure; see also \cite{MA2019} for discussion.
In addition, the connection of group testing with noiseless-coding theory was presented in the group
testing literature by \cite{SG1959} and further investigated in \cite{S1960, S1967}.
In particular, for $N=2$ the procedures $D^{\prime}, S, R_3$, and $R_1$ coincide and are the optimal GT procedures. The optimality follows from the fact that for $N=2$ they are equivalent to the optimal prefix Huffman code \citep{H1952}
with the expected length $L(N)$.
In general, for any $N$,
$L(N)$ can serve as a theoretical lower bound for the expected number of tests of an optimal GT procedure;
however, the complexity of calculation of $L(N)$ is $O\left(2^{N}\log_{2}(2^N)\right)$.
Therefore, even for small $N$, obtaining the exact value of $L(N)$ is impossible.
A well-known noiseless coding theorem provides the information theory bounds for $L(N)$ as
${\displaystyle H(p)\leq L(N) \leq H(p)+1}$, where
${\displaystyle H(p)=N \left[p\log_{2}\frac{1}{p}+q\log_{2}\frac{1}{q}\right]}$
is the Shannon entropy.
For a comprehensive discussion, see \cite{K1973}.

Below, we compare Procedures $D^{\prime}, S$, Optimal Hierarchical ($R_3$), and Optimal Nested ($R_1$) for different $p$
with respect to the expected number of tests for $N=100$. For Procedures $D^{\prime}$ and $S$, the optimal configuration for finite population was found in \cite{G1985} and in \cite{MA2019}, respectively.

\begin{table}[H]
\caption{
The minimal (optimal) expected number of tests per 100 individuals for Procedures $D^{\prime}$, S, Hierarchical ($R_3$), and Nested ($R_1$) for different $p$.
}
\scriptsize
\begin{center}
  \begin{tabular}{lllllll}
     $p$ & $D^{\prime}$ & $S$ & $R_3$ & $R_1$ & $H(p)$\\
     \hline

    0.001 &              6.278 &             4.605& 1.9554&               1.766&         1.141           \\
    0.01 &                19.470&              15.181&9.6872&                8.320&   8.079\\
    0.05 &             41.807&                 36.018&32.0186&               28.958&    28.640\\
    0.10  &                 57.567&                  52.288&50.6752&         47.375&    46.900\\
    0.20 &                  77.872&                 74.974&74.974&       72.875&    72.192\\
    0.25&            84.375 &                 83.875&83.875&           82.191&       81.128\\
    0.30&                    90.500&                    90.500&90.500&             88.889& 88.129\\
    0.35&                  96.375&                 96.375&96.375&               95.633&      93.407\\
    0.38&                  99.780&                  99.780&99.780&                 99.730&      95.804\\
   \bottomrule
  \end{tabular}
  \label{t:1}
  \end{center}
  \end{table}

Table 1 shows that for each value of $p$ there is a consistent ranking among optimal Procedures $D^{\prime}, S$, Optimal Hierarchical ($R_3$), and Optimal Nested ($R_1$) with respect to the expected total number of tests: Procedure $R_1$ is the best, $R_3$ is the second-best, $S$ is the third-best, and $D^{\prime}$ is the worst.
The consistent ranking among Procedures $D^{\prime}, R_3$, and $R_1$ follows from their definitions;
$R_3$ is similar to $D^{\prime}$ but 
without being limited in maximal number of stages as $D^{\prime}$,
and $R_1$ does not have a restriction (as $R_3$ has) that any two units only be tested together in a group if they have an identical test history.
Meanwhile, Procedure $S$ also belongs to the hierarchical class, but has the restriction that in a positive group,
individuals are tested one-by-one until the first positive individual is identified, or until all individuals but the last one are determined negative. Consequently, Procedure $R_3$ and therefore $R_1$ rank higher than Procedure $S$.
To the best of our knowledge, no theoretical results compare optimal Procedures $D^{\prime}$ and $S$.
It is important to note that for some values of $p$ and $N$, ties among the procedures are possible (case $N=2$ was mentioned early in this section).

\section{Nested GT Procedures: heterogeneous $p$ }
The generalized group testing problem (GGTP), first introduced by \cite{S1960}, consists of $N$ stochastically independent units $u_1,u_2,\ldots,u_N$, where unit $u_i$ has the probability $p_i$ ($0<p_i<1$) to be defective and the probability $q_i=1-p_i$ to be non-defective. We assume that the probabilities $p_1, p_2,\ldots,p_N$ are known and that we can decide the order in which the units will be tested. All units have to be classified as either non-defective or defective by group testing.
Since its introduction, GGTP has seen considerable theoretical investigation (\cite{LS1972, NS1973, K1973, H1976a, YH1988a, YH1988b, KS1988, KJP2014, M2017, M2020, MHA2020}).

\bigskip

\noindent
{\bf Dorfman and Sterrett Procedures}
\medskip

\noindent
Ideally, under procedure $A$ ($A\in \left\{D, D^{\prime}, S\right\}$) we are interested in finding an optimal {\it partition} $\displaystyle \left\{m_1,\ldots m_I\right\}$  with $\displaystyle m_1+\ldots+m_I=N$ for some $I\in \left\{1,\ldots,N\right\}$ such that the total expected number of tests is minimal, i.e.
${\displaystyle \left\{m_1,\ldots m_I\right\}=\arg\min_{n_1,\ldots,n_J}
E_{A}\left(n_1,n_2,\ldots,n_J\right)}$ subject to ${ \sum_{i=1}^{J}n_i=N,\,\, J\in\left\{1,\ldots,N\right\}}$,
where  ${\displaystyle E_{A}\left(n_1,n_2,\ldots,n_J\right)=E_{A}\left(1:n_1\right)+\cdots+E_{A}\left(1:n_J\right)}$, and ${\displaystyle E_{A}\left(1:n_j\right)}$ is the total expected number of tests (under procedure $A$) in a group of size $n_j$.  This task is a hard computational problem, and moreover impossible to perform because the total number of possible partitions of a set of size $N$ is the Bell number
$B(N)=\left\lceil \frac{1}{e}\sum_{j=1}^{2N}\frac{j^N}{j\,!}\right\rceil$, which grows exponentially with $N$.
For example, $B(13)=27,644,437$.
In fact, the optimal partition is known only for Procedure $D$, due to \cite{H1975,H1981}.
Hwang proved that under Procedure $D$, an optimal partition is an ordered partition (i.e. each pair of subsets has the property that the numbers in one subset are all greater than or equal to every number in the other subset);
he also provided a dynamic programming algorithm for finding an optimal partition with computational effort $O(N^2)$.
However, the ordered partition is not optimal for Procedures $D^{\prime}$ and $S$ \citep{M2017},
and finding an optimal partition for these procedures is a hard computational problem.
That said, one can evaluate the optimal $D^{\prime}$ and $S$ algorithm under a predetermined order of p's.
In Table 2, we compare Procedures $D, D^{\prime}$ and $S$ for ordered p's, where the method for Procedure $D$ was developed by \cite{H1975} and those for Procedures $D^{\prime}$ and $S$ by \citep{M2017}.
\bigskip

\noindent
{\bf Hierarchical and Nested Procedures}
\smallskip

\noindent
For the fixed predetermined order of $p_1, p_2,\ldots,p_N$ an optimal nested and hierarchical procedures
with respect to the expected total number of tests were developed as DP algorithms in \cite{KS1988} and in \cite{MHA2020}, respectively.
\medskip

\noindent
{\bf Numerical comparisons}
\smallskip

\noindent
We generated the vector $\displaystyle p_1,p_2,\ldots,p_{100}$ from a Beta distribution with parameters $\displaystyle \alpha=1, \beta=(1-p)/p$ such that the expectation equals $p$.
We repeat this process $M=1000$ times for each value of $p$.
Each time an optimal ordered partition with the corresponding expected number of tests was found for Procedures $D$, $D^{\prime}$, and $S$, the hierarchical (HL) and nested (ON) procedures were obtained using DP algorithm (based on the previously mentioned references).
Also, in the GGTP, Shannon entropy $\sum_{i=1}^{N}\left\{p_i\log_{2}\frac{1}{p_i}+(1-p_i)\log_{2}\frac{1}{1-p_i}\right\}$
can serve as the information lower bound for the expected number of tests of an optimal group testing procedure.
The averages of 1000 repetitions are presented in Table 2 below.

\begin{table}[H]
\caption{Comparison of Procedures $D'$, $S$, Hierarchical (HL), and Nested (ON).}
\label{t:1}
\scriptsize
\begin{center}
  \begin{tabular}{llllllll}
    \toprule
    \multirow{1}{*}{p} &
      \multicolumn{5}{c}{$N=100$ } \\
      &{$D^{'}$} & {$S$}& $HL$ &$ON$& {Shannon Entropy}\\
      \midrule
    0.001  &{5.738}&{3.745}&   1.867&1.697&{1.081}           &        \\
    0.01&{17.345}&{13.121}&{8.720}& 7.730&
    {7.474}&\\
    0.05&{37.095}&{31.801}&28.212& 25.797&{25.653}&\\
    0.10&{50.758}&{46.105}& 43.606& 41.192&{40.855}&\\
    0.20&{67.536}&{64.33}&62.69& 61.030&{60.11}&\\
    0.30&{77.598}&{75.358}& 74.382&72.611&{70.303}&\\
   \bottomrule
  \end{tabular}
  \end{center}
  \end{table}

Table 2 shows the same ranking pattern among procedures as was observed in Table 1 for the homogeneous $p$ case. This pattern can be explained along the lines of the previous discussion concerning homogeneous $p$.

\section{Related Issues}
{\bf Unknown $p$}\\
\noindent
In many practical situations, the exact value $p$ of the
probability of disease prevalence is unknown or else only some limited information is available, for example a range.
Since all of the above-presented procedures require knowledge of $p$, there is a need to evaluate $p$
during the process of testing.
For a nested algorithm $R1$, \cite{SG1966} proposed a Bayesian approach that uses upcoming information during testing to evaluate and revaluate $p$. A minimax approach for Procedure $D$ was introduced by \cite{MA2015} and for Procedures
$D^{\prime}$ and $S$ in \cite{MA2019}.
\medskip

\noindent
{\bf{Errors in the Testing}}\\
\noindent
In many settings, particularly in biology and medicine, tests may be subject to measurement error or misclassification. This issue occurs in individual testing but may be enhanced in group testing. In particular, for many applications, the sensitivity of a grouped test may decrease with group size (this is often referred to as dilution).
\cite{GR1972} and \cite{H1976b} recognized early that when tests are misclassified, the objective function should not be the expected number of tests.
\cite{GR1972} and \cite{BM1987} proposed a
modification of the Dorfman procedure and searched for a design that
minimized total cost as a linear function of the expected number of tests,
weighted the expected number of good items misclassified as defective, and
weighted the expected number of defective items misclassified as good. \cite{H1976b}
studied a group testing model with the presence of a dilution effect, where a
group containing a few defective items may be misidentified as one
containing no such items, especially when the size of the group is large. He
calculated the expected cost under the Dorfman procedure in the presence of
the dilution effect and derived the optimal group sizes to minimize this
cost. \cite{MAR2016} characterized the optimal design in the Dorfman
procedure in the presence of misclassification by maximizing
the ratio between the expected number of correct classifications and the
expected number of tests. \cite{HMA2021} proposed to minimize the expected number of tests while controlling overall misclassification rates.
In general, since it is expected that misclassification may be related to group size, one has to be very cautious about proposing  Dorfman designs with large group sizes. Alternative designs where  groups are re-tested in different ways have been explored \citep{LTP1994, LDP2020}.
\medskip

\noindent
{\bf Incomplete identification}\\
\noindent
Consider a very large (infinite) population of items, where each item, independent from the others, is either defective with probability $p$ or non-defective with probability $1-p$. The goal is to identify a certain number of non-defective items as quickly as possible.
To the best of our knowledge, the incomplete identification problem was introduced by \cite{BP1990}.
For recent developments and references, see \cite{M2018}.
\bigskip



{}

\end{document}